# Klessydra-T: Designing Vector Coprocessors for Multi-Threaded Edge-Computing Cores

**Abdallah Cheikh**
Sapienza University of Rome

**Stefano Sordillo**
Sapienza University of Rome

**Antonio Mastrandrea**
Sapienza University of Rome

**Francesco Menichelli**
Sapienza University of Rome

**Giuseppe Scotti**
Sapienza University of Rome

**Mauro Olivieri**
Sapienza University of Rome

**Abstract —** Computation intensive kernels, such as convolutions, matrix multiplication and Fourier transform, are fundamental to edge-computing AI, signal processing and cryptographic applications. Interleaved-Multi-Threading (IMT) processor cores are interesting to pursue energy efficiency and low hardware cost for edge-computing, yet they need hardware acceleration schemes to run heavy computational workloads. Following a vector approach to accelerate computations, this study explores possible alternatives to implement vector coprocessing units in RISC-V cores, showing the synergy between IMT and data-level parallelism in the target workloads.

■ Interleaved multithreading (IMT), or barrel-processing, is a simple and widely known program execution paradigm that alternates instructions belonging to different execution threads in the stages of a single-issue in-order processor pipeline [1,3,4]. In this scheme, while the throughput is limited to 1 instruction per cycle (IPC), pipeline stalls due to inter-instruction dependency are avoided without any hardware overhead for dependency management. As long as the application workload can be programmed as multiple threads, the IMT approach can sustain IPC = 1 with relatively high clock frequency and high energy efficiency, thanks to the hardware simplicity, which is a desirable goal in embedded edge-computing processors.

Nonetheless, to execute computationally heavy applications on the extreme edge, any processor core needs hardware acceleration support. Two broad classes of hardware acceleration exist: hardware units that autonomously execute entire computation kernels upon memory-mapped commands from the processor core, and instruction acceleration units, sometimes referred to as coprocessors, that take over complex instructions and thus are directly sequenced by the core instruction stream. Coprocessors imply less communication overhead, yet they can be efficiently exploited only within Instruction Set Architectures (ISA) that allow extensions dedicated to particular computation domains, such as RISC-V [2].

Edge computing devices regard energy efficiency as the prime concern. This work addresses the introduction of vector coprocessor acceleration in IMT cores for extreme-edge-computing, showing that an IMT processor has an architectural design advantage over other cores with similar IPC, that allows exploiting hardware acceleration with higher energy efficiency and speed.

In this context, we specifically address supporting accelerated vector operations, to execute ubiquitous computation kernels in edge computing applications:
- 2D convolution, covering the broad area of deep neural network applications [6];
- Fast Fourier Transform (FFT), typical of signal processing applications, for example in 5G IoT devices [8];
- Matrix multiplication (MatMul) used in a variety of fields, predominantly in cryptography.

A typical scenario is to run homogenous workloads on all the threads applying the same algorithm on different input data, e.g. convoluting multiple image frames. Otherwise, one can take advantage of the multiple contexts provided in an IMT core and run a composite workload running different algorithms, e.g. transmitting an encrypted stream of a preprocessed video/audio, by convoluting an image while analyzing an audio stream via FFT then encrypting the processed data using an algorithm that heavily relies on MatMul. In this study, we designed, implemented and evaluated a whole taxonomy of coprocessor acceleration schemes for IMT cores, analyzing them for performance, area, and energy efficiency on the above application cases. The contributions of this work are the following:

- We provide designers with a quantitative comparison between different coprocessing schemes referring to different computation kernels;
- Specifically, we identify the optimal balance between Thread Level Parallelism (TLP) and Data Level Parallelism (DLP) in the addressed scenarios;
- We demonstrate the performance and energy efficiency of the IMT approach in the target application contexts by comparing it with processor cores in the same complexity range;
- We show the potentials of an open hardware design based on the RISCV instruction set along with its open programming environment;

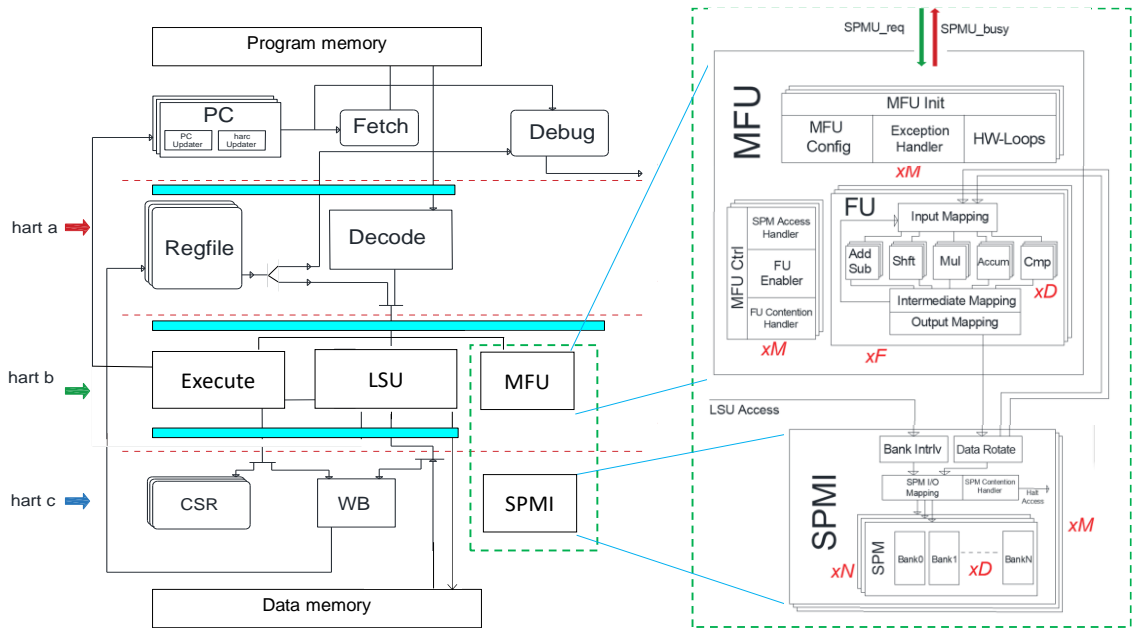

**Figure 1 Klessydra T13 block organization**

BACKGROUND
Many previous works reported the design of hardware accelerated cores in edge-computing applications.
In [14], the authors report the design details of a low-voltage microcontroller with subword-SIMD support.

Our study is more general in investigating various SISD-SIMD-MIMD combinations in coprocessor design. The work in [13] is similar and investigates ad-hoc ISA encoding and pipeline stage balancing for power efficiency and introduces a dedicated coprocessor interface. Yet, the authors do not elaborate on coprocessor architectures and performance. Our work further differs from [13,14] in targeting RISC-V compliance.

In [7], the authors describe a RISC-V processor with DSP hardware support, targeting near-threshold voltage operation, and in the Diet-SODA design [9] a SIMD-oriented DSP accelerator also runs in near-threshold regime. Our study is agnostic about supply or bias voltage tuning, purely addressing DLP and TLP balancing for energy efficiency in any physical implementation, including soft-cores on FPGA, as shown in our results.

A hardware convolution engine for image processing is presented in [12], focusing on the optimal buffer design to store selected portions of the input image. The works in [10, 11] also present convolution accelerators, based on parallel hardware units and local data reuse. Our study adopts a different approach, based on multi-purpose vector coprocessors equipped with scratchpad memories, coupled with an IMT processor, to hide memory latency.

This work builds on the activity reported in [5], that was an initial effort into designing a mathematical accelerator for a RISC-V core, and in [4], that addressed the best performing pipeline organization for an IMT RISC-V core.

## THE KLESSYDRA-T IMT ARCHITECTURE

The processing core discussed in this article, named Klessydra-T13, is a parametric design implementing an IMT four-stage-pipeline RISC-V processor. It supports the RV32IMA instruction set [2], augmented by a custom extension composed of a small subset of mathematical vector instructions. The Klessydra-T13 core (Figure 1) realizes a pure IMT paradigm as defined by the following points:

- Thread context switch at each clock cycle
- in-order, single issue instruction execution
- feed-forward pipeline (no hardware support for branching-hazard and data-hazard handling)
- bare metal execution (RISCV M mode)

The core interleaves three hardware threads (harts [2]) in the instruction pipeline. The register file, program counter, and CSR unit are replicated per hart. A hardware context counter *(harc)* switches between the hart program counters on a rotation basis to fetch instructions from the program memory. The three harts in the four pipeline stages provide a register file access fence, so that it never possible for any two instructions to manifest a dependency hazard in the pipeline.

The T13 core includes multiple units in the execution stage, namely a Load/Store unit (LSU), a scalar execution unit (EXEC) and a vector-oriented multi-purpose functional unit (MFU), which implements the coprocessing features. The LSU works in parallel with other units when executing store instructions, that cannot cause a write-back conflict on the register file.

**Table 1 – Custom vector instruction extension**

| Assembly syntax – (r) denotes memory addressing via register r | Short description |
|---|---|
| `kmemld (rd),(rs1),(rs2)` | load vector into scratchpad region |
| `kmemstr (rd),(rs1),(rs2)` | store vector into main memory |
| `kaddv (rd),(rs1),(rs2)` | adds vectors in scratchpad region |
| `ksubv (rd),(rs1),(rs2)` | subtract vectors in scratchpad region |
| `kvmul (rd),(rs1),(rs2)` | multiply vectors in scratchpad region |
| `kvred (rd),(rs1)` | reduce vector by addition |
| `kdotp (rd),(rs1),(rs2)` | vector dot product into register |
| `ksvaddsc (rd),(rs1),(rs2)` | add vector + scalar into scratchpad |
| `ksvaddrf (rd),(rs1),rs2` | add vector + scalar into register |
| `ksvmulsc (rd),(rs1),(rs2)` | multiply vector + scalar into scratchpad |
| `ksvmulrf (rd),(rs1),rs2` | multiply vector + scalar into register |
| `kdotpps (rd),(rs1),(rs2)` | vector dot product and post scaling |
| `ksrlv (rd),(rs1),rs2` | vector logic shift within scratchpad |
| `ksrav (rd),(rs1),rs2` | vector arithmetic shift within scratchpad |
| `krelu (rd),(rs1)` | vector ReLu within scratchpad |
| `kvslt (rd),(rs1),(rs2)` | compare vectors and create mask vector |
| `ksvslt (rd),(rs1),rs2` | compare vector-scalar and create mask |
| `kvcp (rd),(rs1)` | copy vector within scratchpad region |

The MFU is allowed to read operands from the register file but can write results only to local scratchpad memories (SPMs). The LSU manages data transfers to/from the data memory from/to the SPMs via dedicated instructions.

The MFU executes vector arithmetic instructions, whose latency is proportional to the vector length. A hart requesting access to the busy MFU executes a self-referencing jump until the MFU becomes free, avoiding unnecessary stalls of other harts in the pipeline that are independent from the MFU being busy.

The custom instruction extension supported by the MFU and LSU is summarized in Table 1. The instructions implement vector operations without relying on a vector register file, but rather on a memory space mapped on the local SPMs, for maximum flexibility. The programmer can move vector data at any point of the SPM address space with no constraint except the total capacity of the SPMs, which in turn is a parameter of the microarchitecture design.

The coprocessor instructions are exposed to the programmer as very simple intrinsic functions, fully integrated into the RISC-V *GCC* compiler toolchain.

## HARDWARE ACCELERATION SCHEMES

The MFU and SPMs are accessed through a Scratchpad-Memory Interface (SPMI). The user can configure the number of parallel lanes $D$ in the MFU, the number of MFUs $F$, the SPM capacity, the number of SPMs $N$, the number of SPMIs $M$, and the sharing scheme of MFUs and SPMI among harts. The MFU is the engine that accelerates vector computations. It can operate on different integer data element widths (8, 16, 32-bit) in subword-SIMD fashion, and also in element-SIMD fashion when $D$ is configured to multiply the execution lanes for DLP. A typical vector arithmetic operation has an initial latency between 4 and 8 cycles to access the SPM.

Each SPM has one read and one write port. The parameter $D$ that defines the MFU lanes also corresponds to the number of SPM banks; all the banks of an SPM are accessed together as a single SPM line. When the MFU executes a vector operation, it fetches an entire SPM data line in every clock cycle, composed of multiple vector elements. A bank read rotator aligns the source operands coming from the SPM line, and a bank write rotator aligns the destination data to the correct banks in an SPM line. When the LSU fills the SPM banks with data from the 32-bit data memory port, a bank interleaver switches between the banks. The reader may refer to [5] for internal details of the units inside the MFU and SPMs.

Table 2 – Summary of performance results and synthesis results. Green=best case; Red=worst case.

| Microarchitecture | | | Synthesis results | | | | | | Average Cycle Count per Computation Kernel | | | | | | | | |
|---|---|---|---|---|---|---|---|---|---|---|---|---|---|---|---|---|---|
| | | | FPGA Element Utilization | | | | | Max freq MHz | Homogeneous Workload | | | | | | Composite Workload | | |
| Core | Configuration | DLP | FF | LUT | B-RAM | DSP | LUT-RAM | | Conv 4x4 | Conv 8x8 | Conv 16x16 | Conv 32x32 | FFT 256 | MatMul 64x64 | Conv 32x32 | FFT 256 | MatMul 64x64 |
| Klessydra T13 | SISD | 1 | 2488 | 6982 | 6 | 11 | 264 | 144.4 | 1105 | 3060 | 9727 | 34201 | 33033 | 728187 | 66043 | 80874 | 476771 |
| | SIMD | 2 | 2627 | 8400 | 6 | 15 | 264 | 146 | 895 | 2245 | 6261 | 20374 | 25647 | 602458 | 21976 | 60019 | 645705 |
| | | 4 | 3301 | 11366 | 6 | 23 | 264 | 137.2 | 824 | 1768 | 4607 | 13444 | 22812 | 543164 | 16850 | 29144 | 431773 |
| | | 8 | 4800 | 17331 | 12 | 39 | 264 | 137.7 | 824 | 1613 | 3692 | 10069 | 21555 | 484436 | 11324 | 22482 | 414420 |
| | Sym. MIMD | 1 | 3512 | 10458 | 18 | 19 | 264 | 148.2 | 626 | 1493 | 3887 | 13536 | 18726 | 462066 | 20953 | 17824 | 292564 |
| | Sym. MIMD + SIMD | 2 | 4712 | 15943 | 18 | 31 | 264 | 131.7 | 629 | 1190 | 3123 | 8681 | 16827 | 378748 | 16144 | 15839 | 222370 |
| | | 4 | 6753 | 25089 | 18 | 55 | 264 | 120 | 560 | 1190 | 2543 | 7148 | 15993 | 328962 | 15868 | 14942 | 182580 |
| | | 8 | 10854 | 43419 | 36 | 103 | 264 | 105.1 | 560 | 1152 | 2543 | 6006 | 15726 | 316270 | 15581 | 14613 | 168031 |
| | Het. MIMD | 1 | 3012 | 10182 | 18 | 11 | 264 | 117.2 | 663 | 1521 | 4153 | 13565 | 22839 | 556463 | 27155 | 37111 | 265567 |
| | Het. MIMD + SIMD | 2 | 3871 | 15577 | 18 | 15 | 264 | 128.9 | 638 | 1274 | 3280 | 9167 | 18468 | 425978 | 15973 | 24611 | 251201 |
| | | 4 | 5015 | 23282 | 18 | 23 | 264 | 122 | 573 | 1213 | 2688 | 7473 | 16887 | 360863 | 16042 | 19175 | 181290 |
| | | 8 | 7325 | 42944 | 36 | 39 | 264 | 108.6 | 573 | 1079 | 2580 | 6285 | 17604 | 328178 | 13921 | 17298 | 187877 |
| Klessydra T03 | | | 1418 | 4281 | 0 | 7 | 176 | 221.1 | 1819 | 5737 | 20714 | 79230 | 47256 | 2679304 | 138959 | 46733 | 2775779 |
| RI5CY | | | 2527 | 7674 | 0 | 6 | 0 | 91.4 | 1377 | 4247 | 15088 | 57020 | 37344 | 1360854 | 81534 | 37350 | 1369572 |
| ZeroRiscy | | | 1933 | 5275 | 0 | 1 | 0 | 117.2 | 2510 | 8111 | 29583 | 113793 | 61158 | 4006241 | 197010 | 61163 | 4043376 |

Furthermore, the coprocessor can be configured to implement the following sharing schemes among harts:
**Shared coprocessor:** All the harts share a single MFU/SPM subsystem. In the case of busy MFU, any hart wanting to access it is stalled until the MFU becomes free. In this scheme, parallel execution may occur between coprocessor and non-coprocessor instructions. Yet, the MFU/SPM may exploit pure DLP acceleration, by multi-lane SIMD execution.
**Thread-Dedicated coprocessors:** A complete MFU/SPM subsystem is appointed to each hart, eliminating coprocessor contention. Stalls can only happen if two instructions of the same hart request MFU operation. This scheme can exploit DLP by multi-lane SIMD execution and TLP by fully symmetric MIMD execution, allowing multiple vector instructions to execute in parallel.

**Thread-Dedicated SPMI / Shared MFU:** a dedicated SPM address space is kept for each hart, while the harts share one MFU *at the functional unit level*. This scheme still allows inter-hart parallel execution of coprocessor instructions, provided they use *different* internal functional units of the MFU (e.g. adder, multiplier). Harts requesting a busy internal unit in the MFU get stalled until the contended unit becomes free. This scheme can exploit DLP by multi-lane SIMD execution, and also TLP in the form of a heterogeneous MIMD execution.

The explored design parameters and corresponding configurations, for reference in reporting performance results, are the following:
- $M=1$, $F=1$, $D=1$: *SISD*
- $M=1$, $F=1$, $D=2,4,8$: *Pure SIMD*
- $M=3$, $F=3$, $D=1$: *Symmetric MIMD*
- $M=3$, $F=3$, $D=2,4,8$: *Symmetric MIMD + SIMD*
- $M=3$, $F=1$, $D=1$: *Heterogenous MIMD*
- $M=3$, $F=1$, $D=2,4,8$: *Heterogenous MIMD + SIMD*

We use $N = 3$ in MatMul and $N=4$ in convolutions and FFT.
Finally, we refer to the T13 microarchitecture configured with no hardware acceleration as Klessydra T03.

## PERFORMANCE RESULTS

We run a set of test programs composed of 2D convolution, FFT, and MatMul kernels. We adopted the widely used 3x3 filter size on matrix sizes of 4×4, 8×8, 16×16, and 32×32 elements for convolutions. FFT was run on 256 samples, and MatMul on 64×64 element matrices. The element width was kept 32 bit in fixed-point representation. The tests were organized as follows:

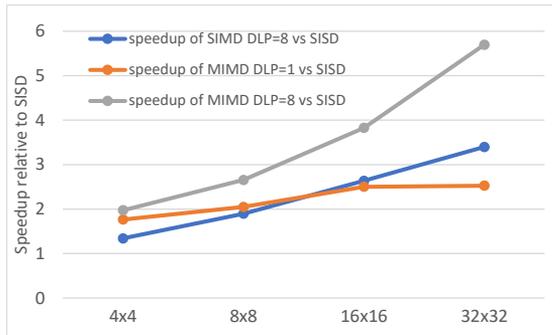

**Figure 2** DLP and TLP cycle-count boost in 2D convolutions for different matrix sizes

- homogeneous workload, running multiple instances of the same kernel on multiple harts, on different input data.
- composite workload, running convolutions, FFTs, and MatMul repeatedly on three respective harts.

The performance was measured by taking the average cycle count to execute one computation kernel. Table 2 summarizes the results, which are discussed below.

**Cycle count:** With small matrix convolutions, the accelerated core reached up to 3× cycle count speed-up over a non-accelerated IMT core (Klessydra T03), and 2× speed-up over single-threaded, DSP-extended core (RI5CY [7]).

As expected, large matrix convolutions and MatMul obtain more considerable advantage from vector-accelerated cores, quantified in 13× cycle count speed-up relative to Klessydra T03, 9× relative to the RI5CY core and 19× relative to ZeroRiscy. In contrast, FFT takes benefit from TLP and reduced data memory accesses rather than from DLP.

Figure 2 quantifies the contribution of DLP and TLP for convolutions on different matrix sizes. For small vectors, TLP inherently exhibits better contributions to speed-up than DLP, while as the vector size grows, the DLP boost dominates. Implementations exploiting both TLP and DLP performed much better than pure DLP also with large matrices. A key outcome is that a single core IMT processor can exploit both DLP and TLP and follow the grey curve, while a single-threaded core exploiting only DLP acceleration follows the blue curve.

Notably, the heterogeneous MIMD coprocessor, that has 3 times less functional units than the fully symmetric MIMD, employed only 1% to 7% more cycles than the latter.

**Maximum clock frequency:** All the cores under analysis were implemented as FPGA soft-cores. The clock speed exhibited the sharpest drops as the TLP grew larger: in the heterogeneous MIMD scheme, the crossbar mapping the SPMI output data on the shared MFU units became the critical path for D=4,8. Pipelining the crossbar to reduce the critical path, introduces hardware overhead, compromising the area advantage of the heterogeneous MIMD configuration.

**Absolute execution time:** The cycle count and the operating frequency allow calculating the total execution time. Figure 3 compares the actual execution time speed-up relative to the ZeroRiscy core, taken as the reference when each core operates at its maximum frequency. In pure SIMD configurations, the speed-up grows linearly with the DLP for the explored DLP range. Yet, exploiting TLP, by going from a SISD/SIMD to symmetric and heterogenous MIMD, improved the speedup in all cases, despite the frequency drop associated with the MIMD coprocessor. Thanks to exploiting both TLP and DLP, the *symmetric MIMD+SIMD* schemes exhibit the lowest execution times, reaching up to 17× speed-up over Zeroriscy for Convolution 32x32 and up to 13× speed-up for the composite workload. Notably, the *heterogeneous MIMD* configurations maintain an almost perfect overlap with the *symmetric MIMD*.

The non-accelerated Klessydra-T03, while employing a higher cycle count than RI5CY due to the absence of DSP and hardware-loop extensions, exhibits an absolute performance advantage over RI5CY thanks to a more than double frequency attained by the pure IMT microarchitecture. When compared to ZeroRiscy, T03 exhibits both lower cycle count and higher frequency.

**Hardware Resource Utilization:** In cost-constrained applications, it is crucial to find an optimal balance between speed-up and area overhead. The *heterogenous MIMD + SIMD* scheme with $D = 2$ resulted to be a possible best choice with all test programs.

The non-accelerated T03 exhibits only a slightly more significant footprint than the tiny ZeroRiscy core, despite the replicated register file to support multi-threading, thanks to the LUT-RAM implementation of the registers.

**Energy Efficiency:** The average energy per algorithmic operation (multiplications and additions) is a general measure of the energy efficiency attained by a processor core in implementing an algorithm computation. Figure 4 reports the outcome of this analysis, referring to the soft-core implementations. The results are presented as the reduction in nJ/op relative to Zeroriscy, taken as reference, which exhibited 4.24 nJ/op as the best case in the analyzed workloads.

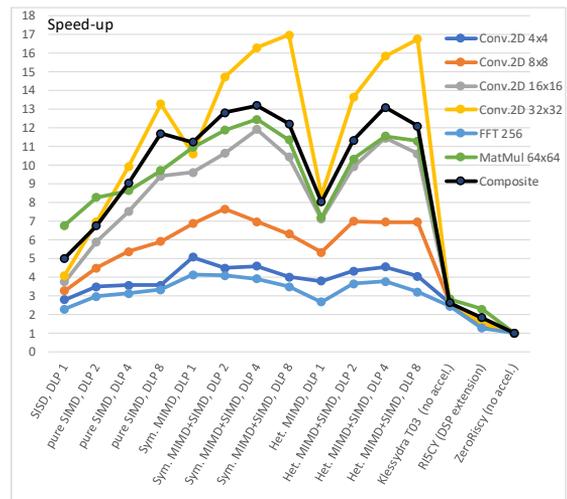

**Figure 3** Execution time speed-up with respect to Zeroriscy core, taken as reference. For the composite test the average kernel speed-up is reported.

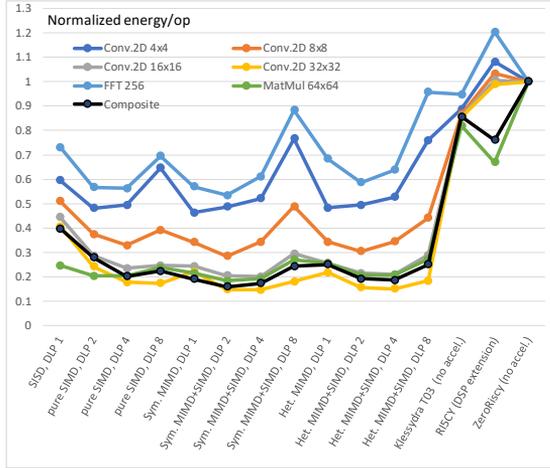

**Figure 4** Average energy per algorithmic operation, normalized to the case of the Zeroriscy soft-core, taken as reference.

The most energy efficient designs resulted to be the *symmetric MIMD* and *heterogenous MIMD* schemes, again exhibiting an almost complete overlap and reaching over 85% energy saving related to the reference Zeroriscy. Despite having the smallest area footprint, the pure SIMD schemes resulted in a larger energy consumption, due to low exploitation of TLP.

**Larger Filters:** convolutional neural networks primarily employ 3×3 filters (VGG16) but also larger ones (e.g. 11×11 in Alexnet, 5×5 in Googlenet). Large masks such as 7×7 are used in Sobel, Gaussian smoothing, median filtering. We evaluated the vector coprocessor schemes with filters ranging from 5×5 to 11×11, on 32×32 element matrices. Table 3 shows the speed-up and energy efficiency trends continue as the filter dimensions grow larger, favoring higher DLP. The improvement referring to ZeroRiscy grows up to 15× when using 11×11 filters.

The *symmetric* and *heterogeneous MIMD+SIMD* schemes, with $D=2$, maintain similar performance and energy results throughout the analyzed cases. The results confirm that an IMT core capable of MIMD acceleration increasingly performs better than a single-thread SIMD acceleration.

CONCLUSIONS

The scientific outcome of this study can be summarized in the following list of evidence:

- The *MIMD-SIMD* vector coprocessor schemes enable tuning the TLP and DLP contribution and obtain the best results in absolute performance and energy efficiency, reaching >15× speed-up and -85% energy per operation.

- Kernels that are less effectively vectorizable can still benefit from acceleration through SPMs and TLP, in an IMT core, reaching 2×-3× speed-up.

- Fully symmetric and heterogeneous MIMD give very similar results, showing that coprocessor contention can be effectively mitigated by functional unit heterogeneity, allowing hardware resource saving. From the same observation, we can state that functional unit contention is less impacting than SPM contention, in all the kernels.

- Pure DLP acceleration always gives inferior results than a balanced TLP/DLP acceleration. An IMT microarchitecture can benefit from TLP and DLP acceleration in a single core.

- In the absence of hardware acceleration, IMT still exhibits an absolute performance advantage over

**Table 3** Higher order filter evaluation results for cycle count, total time at max frequency and total energy. Green=best case; Red=worst case.

| Core | DLP | Filter (5x5) | | | Filter (7x7) | | | Filter (9x9) | | | Filter (11x11) | | |
|---|---|---|---|---|---|---|---|---|---|---|---|---|---|
| | | Cycle Cnt ×1000 | T [us] | E [uJ] | Cycle Cnt ×1000 | T [us] | E [uJ] | Cycle Cnt ×1000 | T [us] | E [uJ] | Cycle Cnt ×1000 | T [us] | E [uJ] |
| T13 SIMD | 2 | 53 | 362 | 51 | 101 | 694 | 97 | 166 | 1136 | 159 | 247 | 1689 | 237 |
| T13 SIMD | 8 | 25 | 179 | 34 | 46 | 335 | 65 | 75 | 543 | 105 | 111 | 803 | 155 |
| T13 Sym MIMD | 2 | 20 | 148 | 27 | 36 | 272 | 49 | 57 | 436 | 79 | 84 | 641 | 117 |
| T13 Sym MIMD | 8 | 12 | 113 | 29 | 19 | 183 | 47 | 30 | 284 | 73 | 43 | 408 | 105 |
| T13 Het MIMD | 2 | 21 | 159 | 28 | 38 | 291 | 52 | 60 | 467 | 83 | 89 | 687 | 122 |
| T03 (no accel.) | - | 247 | 1120 | 216 | 515 | 2328 | 448 | 881 | 3985 | 767 | 1369 | 6191 | 1191 |
| RISCY | - | 180 | 1971 | 252 | 385 | 4218 | 539 | 663 | 7252 | 928 | 1000 | 10949 | 1400 |
| ZeroRiscy | - | 319 | 2721 | 226 | 675 | 5754 | 479 | 1130 | 9637 | 802 | 1698 | 14482 | 1205 |

single-thread execution thanks to the simplified hardware structure.

The Klessydra-T parametric cores are available as open source designs on GitHub at https://perma.cc/6FYD-AF68 .

## ■ REFERENCES